\documentclass[lettersize,journal]{IEEEtran}

\usepackage[utf8]{inputenc}

\AtBeginDocument{
  \providecommand\BibTeX{{
    Bib\TeX}}}
    
\usepackage{amsmath,amsfonts}
\usepackage{authblk}
\usepackage{algorithmic}
\usepackage{array}
\usepackage[caption=false,font=normalsize,labelfont=sf,textfont=sf]{subfig}
\usepackage{textcomp}
\usepackage{stfloats}
\usepackage{url}
\usepackage{verbatim}
\usepackage{graphicx}
\usepackage{booktabs}
\usepackage[per-mode=symbol,retain-explicit-plus,separate-uncertainty=true,detect-all=true]{siunitx}
\def\defaultDigitGrouping{5}
\def\BibTeX{{\rm B\kern-.05em{\sc i\kern-.025em b}\kern-.08em
    T\kern-.1667em\lower.7ex\hbox{E}\kern-.125emX}}
\usepackage{balance}

\usepackage{graphicx}
\usepackage{svg}
\usepackage[per-mode=symbol]{siunitx}
\usepackage{stfloats}
\usepackage{titlesec}
\usepackage{pifont}
\usepackage{orcidlink}
\usepackage{siunitx}
\DeclareSIUnit{\cycle}{cycle}

\newcommand{\cmark}{\ding{51}} 
\newcommand{\xmark}{\ding{55}} 

\definecolor{mygreen}{HTML}{007480}
\definecolor{myred}{HTML}{6D1A36}

\newcommand{\orcid}[1]{\href{https://orcid.org/#1}{\includegraphics*[width=8pt]{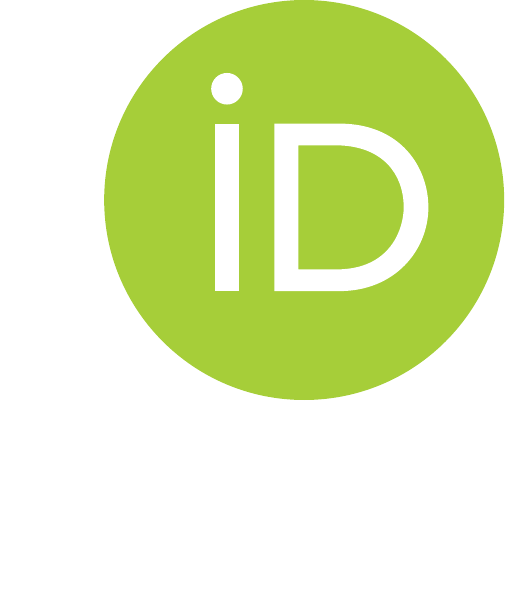}}}

\begin{document}

\title{e-GPU: An Open-Source and Configurable RISC-V Graphic Processing Unit for TinyAI Applications}

\author{Simone~Machetti\orcid{0000-0002-2887-5031},
        Pasquale~Davide~Schiavone\orcid{0000-0003-2931-0435},
        Lara~Orlandic\orcid{0000-0002-4078-7528},
        Darong~Huang\orcid{0000-0002-6579-0627},
        Deniz~Kasap\orcid{0000-0002-6318-1705},
        Giovanni~Ansaloni\orcid{0000-0002-8940-3775},
        David~Atienza\orcid{0000-0001-9536-4947},~\IEEEmembership{Fellow,~IEEE}
\thanks{All authors are with the Embedded Systems Laboratory (ESL), EPFL, Lausanne, Switzerland. Email: simone.machetti@epfl.ch.}}

\maketitle

\begin{abstract}

    Graphics processing units (GPUs) excel at parallel processing, but remain largely unexplored in ultra-low-power edge devices (TinyAI) due to their power and area limitations, as well as the lack of suitable programming frameworks. To address these challenges, this work introduces embedded GPU (e-GPU), an open-source and configurable RISC-V GPU platform designed for TinyAI devices. Its extensive configurability enables area and power optimization, while a dedicated Tiny-OpenCL implementation provides a lightweight programming framework tailored to resource-constrained environments. To demonstrate its adaptability in real-world scenarios, we integrate the e-GPU with the eXtendible Heterogeneous Energy-Efficient Platform (X-HEEP) to realize an accelerated processing unit (APU) for TinyAI applications. Multiple instances of the proposed system, featuring varying e-GPU configurations, are implemented in TSMC’s \SI{16}{\nano\meter} SVT CMOS technology and are operated at \SI{300}{\mega\hertz} and \SI{0.8}{\volt}. Their area and leakage characteristics are analyzed to ensure alignment with TinyAI constraints. To assess both runtime overheads and computational efficiency, we employ two benchmarks: General Matrix Multiply (GeMM) and bio-signal processing (TinyBio) workloads. The GeMM benchmark is used to quantify the scheduling overhead introduced by the Tiny-OpenCL framework. The results show that the delay becomes negligible for matrix sizes larger than $256 \times 256$ (or equivalent problem sizes). The TinyBio benchmark is then used to evaluate performance and energy improvements over the baseline host under pure processing conditions. The results indicate that the high-range e-GPU configuration with 16 threads achieves up to a \SI{15.1}{\times} speed-up and reduces energy consumption by up to \SI{3.1}{\times}, while incurring only a \SI{2.5}{\times} area overhead and operating within a \SI{28}{\milli\watt} power budget.

\end{abstract}

\begin{IEEEkeywords}

    Ultra-Low-Power, Open-Source, Graphic Processing Unit, Microcontroller, Artificial Intelligence.
    
\end{IEEEkeywords}


\section{INTRODUCTION}

    The growing need for machine learning-based real-time computing has fueled the rapid expansion of edge computing. By processing data locally rather than relying on cloud servers, edge computing reduces latency, enhances privacy, and improves energy efficiency, making it an ideal solution for numerous edge applications. However, these workloads impose limitations on computational performance, real-time responsiveness, and power consumption, necessitating specialized hardware architectures~\cite{co-design_vision}.
    
    A promising approach to meeting these challenges is the adoption of heterogeneous architectures, which integrate a host central processing unit (CPU) with domain-specific accelerators to balance efficiency and performance~\cite{heepocrates, asip}. Among these accelerators, graphics processing units (GPUs) have proven particularly effective in exploiting data parallelism for machine learning and signal processing tasks. When coupled with host CPUs, GPUs form accelerated processing units (APUs), enabling a unified platform that efficiently handles both general-purpose tasks and computationally intensive workloads. 

    While various GPU implementations exist, ranging from high-performance to embedded solutions, their trade-offs in the context of ultra-low-power edge devices (TinyAI) remain largely unexplored. These battery-powered devices operate under tight power constraints, typically in the range of tens of milliwatts, necessitating highly efficient GPU architectures. Their small form factor also imposes strict area restrictions, with a complete system-on-chip (SoC) occupying only a few square millimeters. Moreover, the absence of a file system and multi-threading support prevents the use of traditional GPU programming frameworks, such as standard open computing language (OpenCL) implementations~\cite{pocl}, necessitating custom optimizations.

    Due to the aforementioned limitations, ultra-low-power devices typically rely on specialized accelerators such as coarse-grained reconfigurable arrays (CGRAs)~\cite{vwr2a, vwr2a_2}, systolic arrays~\cite{cnn}, and near-memory~\cite{carus} or in-memory computing~\cite{blade} solutions, while the potential of GPUs in this domain remains largely unexplored.

    This work examines the feasibility and trade-offs associated with utilizing GPUs for TinyAI applications by introducing an open-source and configurable RISC-V platform, embedded GPU (e-GPU). The extensive configurability of the platform enables area and power optimization to meet the requirements of this domain. Furthermore, a custom tiny open computing language (Tiny-OpenCL) implementation overcomes the aforementioned limitations and provides a lightweight programming framework specifically developed for resource-constrained devices. To demonstrate the adaptability of the proposed platform in real-world scenarios, we integrate the e-GPU with the eXtendible Heterogeneous Energy-Efficient Platform (X-HEEP)~\cite{x-heep, x-heep_1} to realize an APU for TinyAI workloads. 
    
    Multiple instances of the proposed system, featuring varying e-GPU configurations, are implemented in TSMC’s \SI{16}{\nano\meter} SVT CMOS technology and are operated at \SI{300}{\mega\hertz} and \SI{0.8}{\volt}. Their area and leakage characteristics are analyzed to ensure alignment with TinyAI constraints. To assess both runtime overheads and application-level efficiency, we employ two benchmarks: General Matrix Multiply (GeMM) and bio-signal processing (TinyBio) workloads. The GeMM benchmark is used to quantify the scheduling overhead introduced by the Tiny-OpenCL framework, while the TinyBio benchmark is then used to evaluate performance and energy improvements over the baseline host.
    
    The key contributions of this work are as follows:
    
    \begin{itemize}
        \item We advocate domain-specific GPUs as a suitable solution for the TinyAI domain.
        \item We present e-GPU~\cite{e-gpu}, an open-source and configurable RISC-V GPU platform designed for TinyAI workloads.
        \item We analyze the programming limitations of this application domain and introduce a Tiny-OpenCL framework for resource-constrained devices.
        \item We explore the feasibility and trade-offs of utilizing GPUs for TinyAI applications.
        \item We release an open-source repository, including the complete e-GPU\footnote{The complete code and documentation of the e-GPU platform is \mbox{open-source} on GitHub.} to allow researchers to tailor the platform to their TinyAI domains.
    \end{itemize}

    The remainder of this paper is organized as follows. Section~2 discusses related works. Section~3 summarizes background concepts. Section~4 describes the e-GPU hardware, while Section~5 focuses on the e-GPU software. Section~6 explains the integration with a host. Section~7 outlines the experimental setup, and Section~8 presents the experimental results. Finally, Section~9 concludes the paper.

\section{STATE-OF-THE-ART}

    This section analyzes the most relevant commercial and academic GPU solutions that have informed our platform's design choices.

    \subsection{Commercial GPUs}

        The landscape of commercial edge GPUs is diverse among solutions that balance performance and power efficiency. Qualcomm’s Adreno GPUs, integrated into Snapdragon SoCs, enable power-efficient computing for mobile and edge applications. The Adreno 600 series, in particular, is widely used for machine learning inference, using advanced energy-saving mechanisms. Similarly, ARM’s Mali GPUs, embedded in various SoCs, have evolved with power-optimized architectures. Innovations such as fine-grained power gating in the Mali-G52 and Mali-G72 series improve energy efficiency, making them suitable for low-power scenarios. Another key player, the PowerVR GPUs from Imagination Technologies, prioritizes power-efficient performance, with the Series8XT demonstrating competitive computing capabilities in edge devices.

        Despite these advancements, commercial GPUs are not explicitly designed for TinyAI applications. Their power consumption, typically ranging from hundreds of milliwatts to several watts, exceeds the stringent requirements of these applications, which demand power levels in the order of tens of milliwatts. Furthermore, the proprietary nature of these GPUs prevents detailed performance and power characterization, limiting their use in research and energy-efficient design exploration.

    \subsection{Academic GPUs}

        Academic GPU research has focused on developing programmable and configurable architectures for various computing domains. Table~\ref{tbl:gpu_table} compares GPU platforms based on six essential features: (1)~open-source code for quick access, (2)~open-source instruction-set architecture (ISA) for easy custom extensions, (3)~configurability to adapt to specific application needs, (4)~synthesizable register transfer level (RTL) code for accurate performance and power evaluation, (5)~integration capability with a host CPU to study CPU-GPU interactions, and (6)~ultra-low-power design for TinyAI applications.

        Academic GPU solutions range from software simulators to hardware implementations.

        \subsubsection{Software Simulators} 
            
            GPGPU-Sim~\cite{GPGPU-sim} models CUDA-based GPUs, emulating architectures such as NVIDIA Fermi and GT200. gem5-gpu~\cite{gem5-gpu} extends this by integrating the x86 CPU and memory models of gem5~\cite{gem5}, facilitating studies on CPU-GPU interactions. Another example, Multi2Sim~\cite{multi2sim}, simulates the AMD Evergreen GPU family alongside multi-threaded x86 processors.

            While software simulators facilitate rapid architectural exploration, they have significant limitations for TinyAI research, including the absence of synthesizable RTL code, which prevents accurate power and energy analysis, the lack of open-source ISA, restricting opportunities for extensions, and their primary focus on high-performance computing, making them unsuitable for ultra-low-power edge applications.

        \subsubsection{Hardware Implementations}
                
            Several academic GPU projects provide synthesizable RTL code, offering deeper insights into power and performance metrics. Skybox~\cite{skybox} is a RISC-V GPU framework designed for graphics research with Vulkan API support. However, its reliance on PCIe-based communication with a host PC makes it unsuitable for integration into SoCs, which are commonly used in ultra-low-power applications. ZJX-RGPU~\cite{zjx-rgpu} integrates a \mbox{RISC-V} CPU and a fragment engine into a \SI{12.25}{\milli\meter\squared} chip, operating at \SI{200}{\mega\hertz} with a power consumption of approximately \SI{20}{\milli\watt}, but its closed-source nature and limited documentation restrict its use for architectural exploration.
            
            Vortex~\cite{vortex} represents a significant advance as a configurable RISC-V GPU that supports OpenCL and OpenGL. Synthesized with a single compute unit using a \SI{15}{\nano\meter} educational cell library, it consumes \SI{46.8}{\milli\watt} at \SI{300}{\mega\hertz}. However, its hardware is not natively synthesizable without extensive modification, and it targets PC-class host environments. These characteristics hinder its integration into SoCs, where lightweight interconnects and tight hardware-software coupling are essential for ultra-low-power applications.
            
            The METASAT platform~\cite{metasat} adapts Vortex for a safety-critical RISC-V SoC, integrating it with the NOEL-V processor and the SPARROW single-instruction multiple-data (SIMD) accelerator. Designed to run under RTEMS and the XtratuM hypervisor, the platform removes OpenCL dependencies and enables GPU usage in partitioned real-time environments. However, its GPU inherits Vortex's limitations and serves only as a proof-of-concept for integrating GPUs into future space systems rather than a solution for TinyAI computing, where energy efficiency is the primary concern.

            Virgo~\cite{virgo}, built on Vortex, introduces a disaggregated GPU architecture with cluster-level matrix units, improving data reuse and reducing energy consumption compared to core-coupled designs. While Virgo achieves high energy efficiency and is fully open-source, it targets large-scale GEMM operations and datacenter-class deep learning workloads, limiting its flexibility for TinyAI scenarios with smaller problem sizes. Additionally, Virgo relies on a relatively complex memory hierarchy and a sophisticated synchronization mechanism across compute units, which may be challenging to adapt to minimalist SoC designs with tight area and power constraints.
            
            Ventus~\cite{ventus} builds on the RISC-V vector extension (RVV) and introduces a high-performance GPGPU architecture with custom instructions for single-instruction multiple-thread (SIMT) execution, predicated branching, synchronization, and tensor operations. It scales up to 16 compute units and 256 warps, demonstrating up to \SI{87.4}{\percent} CPI and \SI{83.9}{\percent} instruction count reductions over Vortex. However, Ventus is designed exclusively for deployment on large-scale FPGAs and has not been integrated into an SoC, targeting PC-class hosts. Although synthesis results are reported using a TSMC \SI{12}{\nano\meter} library for frequency and area evaluation, no low-power or tape-out optimizations are presented. As such, Ventus targets high-throughput applications and architectural research rather than TinyAI scenarios.
            
            \begin{table*}[tp]
                \caption{Comparison of relevant academic GPU platforms based on key features.}
                \label{tbl:gpu_table}
                \centering
                \resizebox{0.995\textwidth}{!}{
                \begin{tabular}{|l|c|c|c|c|c|c|}
                \toprule
                \bf{GPU Platform} & \bf{Open-source} & \bf{Open ISA} & \bf{Configurable} & \bf{Synthesizable} & \bf{Integrable} & \bf{Ultra-low-power} \\
                \midrule
                GPGPU-Sim~\cite{GPGPU-sim} & \textcolor{mygreen}{\cmark} & \textcolor{myred}{\xmark} & \textcolor{mygreen}{\cmark} & \textcolor{myred}{\xmark} & \textcolor{myred}{\xmark} & \textcolor{myred}{\xmark}\\
                gem5-gpu~\cite{gem5-gpu} & \textcolor{mygreen}{\cmark} & \textcolor{myred}{\xmark} & \textcolor{mygreen}{\cmark} & \textcolor{myred}{\xmark} & \textcolor{mygreen}{\cmark} & \textcolor{myred}{\xmark}\\
                Multi2Sim~\cite{multi2sim} & \textcolor{mygreen}{\cmark} & \textcolor{myred}{\xmark} & \textcolor{mygreen}{\cmark} & \textcolor{myred}{\xmark} & \textcolor{mygreen}{\cmark} & \textcolor{myred}{\xmark}\\
                \midrule
                Skybox~\cite{skybox} & \textcolor{mygreen}{\cmark} & \textcolor{mygreen}{\cmark} & \textcolor{mygreen}{\cmark} & \textcolor{mygreen}{\cmark} & \textcolor{myred}{\xmark} & \textcolor{myred}{\xmark}\\
                ZJX-RGPU~\cite{zjx-rgpu} & \textcolor{myred}{\xmark} & \textcolor{mygreen}{\cmark} & \textcolor{myred}{\xmark} & \textcolor{mygreen}{\cmark} & \textcolor{mygreen}{\cmark} & \textcolor{mygreen}{\cmark}\\
                Vortex~\cite{vortex} & \textcolor{mygreen}{\cmark} & \textcolor{mygreen}{\cmark} & \textcolor{mygreen}{\cmark} & \textcolor{myred}{\xmark} & \textcolor{myred}{\xmark} & \textcolor{myred}{\xmark}\\
                METASAT~\cite{metasat} & \textcolor{mygreen}{\cmark} & \textcolor{mygreen}{\cmark} & \textcolor{mygreen}{\cmark} & \textcolor{myred}{\xmark} & \textcolor{mygreen}{\cmark} & \textcolor{myred}{\xmark}\\
                Virgo~\cite{virgo} & \textcolor{mygreen}{\cmark} & \textcolor{mygreen}{\cmark} & \textcolor{mygreen}{\cmark} & \textcolor{mygreen}{\cmark} & \textcolor{mygreen}{\cmark} & \textcolor{myred}{\xmark}\\
                Ventus~\cite{ventus} & \textcolor{mygreen}{\cmark} & \textcolor{mygreen}{\cmark} & \textcolor{mygreen}{\cmark} & \textcolor{mygreen}{\cmark} & \textcolor{myred}{\xmark} & \textcolor{myred}{\xmark}\\
                \midrule
                \textbf{Proposed e-GPU} & \textcolor{mygreen}{\cmark} & \textcolor{mygreen}{\cmark} & \textcolor{mygreen}{\cmark} & \textcolor{mygreen}{\cmark} & \textcolor{mygreen}{\cmark} & \textcolor{mygreen}{\cmark}\\
                \bottomrule
                \end{tabular}
                }
            \end{table*}

\section{Background}

    This section summarizes the terminology needed on the basis of the standard OpenCL definitions. The terms are organized according to the platform, execution, and runtime models. The platform model defines the relationship between the host system and the compute devices responsible for executing kernels. The execution model describes how kernels are launched and how work-items and work-groups are organized and executed across the available hardware. Finally, the runtime model specifies the behavior of kernel execution at the hardware level, including thread scheduling, synchronization, and utilization of execution resources.

    \subsection{Platform Model}
    
        In the platform model, the host refers to the main system, typically a CPU, that runs the OpenCL host code and is responsible for managing and dispatching kernel execution across compute devices. A device is an accelerator, such as a GPU, that executes OpenCL kernels. Each device is composed of one or more compute units, which are the architectural blocks responsible for executing work-groups. Within a compute unit are multiple processing elements, which are the lowest-level execution resources and carry out the arithmetic and logic operations of the work-items.
    
    \subsection{Execution Model}
    
        The execution model governs how computation is organized and launched. The host code orchestrates the execution process from the host side, issuing commands to devices and managing data transfers. A kernel is an OpenCL function that runs on the compute device and is executed in parallel by many work-items. Each work-item corresponds to a single instance of kernel execution and operates on a distinct portion of the input data, identified by a unique global ID. Work-items are grouped into work-groups, which are assigned to compute units. Work-items within a group can share data through local memory and synchronize with one another using barriers. The global size specifies the total number of work-items to be launched, while the local size determines how many of these are grouped into each work-group.
    
    \subsection{Runtime Model}
    
        The runtime model describes how kernels are executed at the hardware level. A thread represents one instance of a kernel running on a processing element, maintaining its own private set of registers. Threads are often grouped into warps, which are sets of threads that execute in lockstep, meaning they share the same program counter and execute the same instruction simultaneously. Depending on hardware capabilities, multiple warps may be active concurrently in the pipeline, enabling fine-grained parallelism and efficient utilization of execution resources.

\section{{\textnormal{e}}-GPU HARDWARE}

    This section describes the e-GPU architecture and the software-hardware co-design decisions aimed at minimizing area, power, and energy consumption while adapting the platform to the TinyAI domain. Figure~\ref{fig:gpu_architecture} illustrates the architecture of the proposed e-GPU, while Table~\ref{tbl:gpu_knobs} lists the available configurable knobs, which are accessible through the SystemVerilog parameters.

    \subsection{Compute Unit} \label{sec:CU}

        The compute unit is based on the Vortex core~\cite{vortex}. It features an out-of-order RISC-V architecture with internal configurability options that allow tuning parallel threads and concurrent warps. Adjusting the number of threads modifies data parallelism by varying the number of processing elements, while configuring the number of warps enhances resource and memory bandwidth utilization by enabling multiple warps to execute concurrently through fine-grained multi-threading. 

        Due to power and area limitations, we removed the floating-point unit to prioritize integer and fixed-point arithmetic operations, as the platform targets TinyAI models based on integer numbers. Furthermore, we extended the RISC-V ISA with a custom \textsc{SLEEP\_REQ} instruction to improve energy efficiency. At the end of the kernel processing, this instruction waits for all previously fetched instructions to execute before generating an event to the GPU controller, signaling the completion of the operations. This mechanism enables the GPU controller to power down the compute unit through clock or power gating, if available, as soon as it is no longer in use, thereby improving the overall energy efficiency. At design time, the number of compute units can be configured to trade area, power, and processing requirements.

    \subsection{Memory Hierarchy}

        The memory hierarchy adopts a unified architecture in which the host main memory and the e-GPU global memory map to the same physical memory on the host. This shared-memory model enhances programmability by eliminating the need for explicit data transfers between separate memory regions. In particular, complex explicit dual-buffering schemes are no longer required to move input and output data between the host main memory and the e-GPU global memory. 

        Each compute unit includes a private instruction cache to reduce memory access latency and improve performance, while a shared data cache is used across compute units.
        
        The cache logic is based on the Vortex cache~\cite{vortex} and features a direct-mapped, multi-bank architecture with a line-interleaved addressing scheme. Support for outstanding requests allows subsequent warps to issue new cache accesses even if a previous warp is stalled due to a cache miss. At design time, the cache size, the number of banks, and the line size can be adjusted based on the specific requirements of the target application. We integrated dedicated memory wrappers to instantiate production-level SRAM macros for each instruction bank seamlessly.

        Each compute unit is connected to a private instruction cache. When multiple compute units execute the same kernel, this design improves scalability as the number of compute units increases and avoids conflicts that would otherwise arise with a shared instruction cache. In fact, there is no guarantee that each compute unit fetches the same instruction at each clock cycle. Thus, having a separate private instruction cache helps reduce misses and mitigates performance degradation. Furthermore, separate caches improve locality when different compute units execute different kernels.

        On the data side, all compute units share a common data cache. This design provides an intermediate storage layer between the slower main memory of the host and the faster register files of the compute units, thus improving the efficiency of data access. Additionally, using a shared cache enables direct data sharing between compute units without requiring additional transfers to and from the host main memory.

        Input data is loaded from the host into the shared data cache at the beginning of the first kernel and remains available throughout the execution of intermediate kernels, assuming that the cache is large enough to hold the entire working set. This approach improves performance by minimizing data transfers, which ultimately results in improved energy efficiency.

        Finally, we redesigned the data cache interface to support multi-threaded requests from all compute units in parallel. Each compute unit issues multiple simultaneous requests, one per active thread, but receives a single unified response. Input and output data are selectively masked on the basis of the requesting threads. This support for multidimensional requests enhances bandwidth utilization by enabling concurrent access to the shared cache when no bank conflicts occur.

        At the system boundary, the memory interface efficiently adapts cache requests to host memory transactions. These are managed in three steps. (1)~On a miss, the cache requests an entire multi-word line, which is then serialized into individual 32-bit transactions. (2)~The cache protocol is translated into the OBI~\cite{obi} protocol, a low-power bus protocol widely adopted in edge systems. (3)~An arbiter~\cite{bus} manages access to the shared output port, ensuring fair and efficient resource allocation.

        \begin{figure*} [t]
            \centering
            \includegraphics[width=0.99\textwidth]{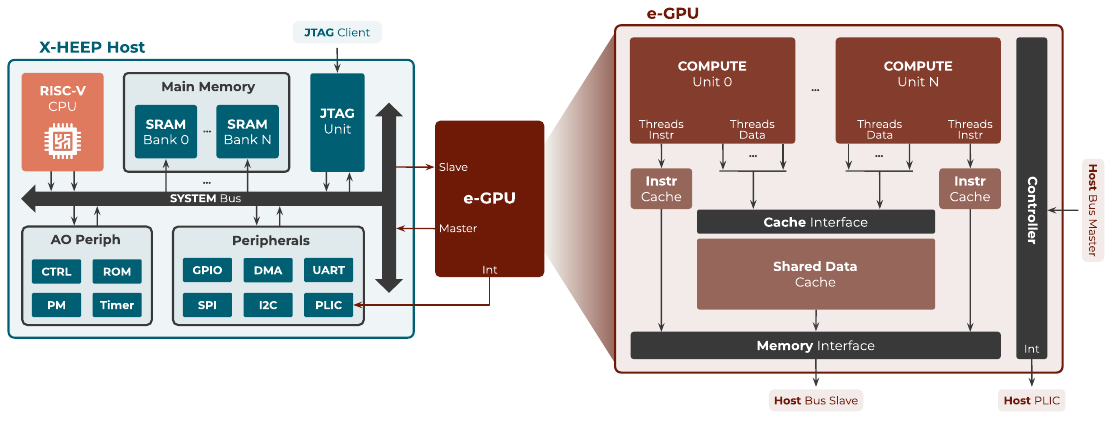}
            \caption{Architecture of the presented e-GPU platform and its integration with the X-HEEP host~\cite{x-heep}.}
            \label{fig:gpu_architecture}
        \end{figure*}

    \subsection{Controller}

        We designed a dedicated controller to manage the operations of the e-GPU. A set of memory-mapped configuration registers enables operations such as resetting, starting, and halting the accelerator. Furthermore, the controller integrates a power controller that operates in conjunction with the low-power RISC-V extension described in Section~\ref{sec:CU}.

        After a kernel is launched, the power controller monitors end-of-execution events from each compute unit. As soon as an event is received, the corresponding compute unit is powered down. Once all events have been received, indicating that all compute units have completed execution, the controller generates an external interrupt to notify the host CPU.

\section{{\textnormal{e}}-GPU SOFTWARE} \label{sec:opencl}

    This section describes the Tiny-OpenCL framework, a lightweight and highly optimized OpenCL implementation designed to compile kernels for execution on the e-GPU. The framework maximizes compatibility with existing GPU software while ensuring portability across platforms.

    \subsection{Tiny-OpenCL framework}
    
        The framework includes the following software components: (1)~the SIMT RISC-V extension API~\cite{vortex}, which provides access to low-level compute unit functionalities, such as activating and deactivating threads and warps, handling divergences through splitting and joining, and managing synchronization barriers; (2)~the startup functions, responsible for initializing the e-GPU and performing essential setup operations, including computing the stack pointer for each thread; and (3)~the scheduler functions, which orchestrate the execution of work-items across the available compute resources (compute units, threads, and warps), ensuring efficient workload distribution. These components are precompiled into a static library using the standard RISC-V GNU toolchain.
    
        Subsequently, the OpenCL kernel is analyzed by a parser script, which performs the necessary code transformations to translate it into a standard C function with equivalent functionality. This C function can then be compiled using the RISC-V GNU toolchain to produce the corresponding object file. These transformations enable seamless integration with the existing compilation tools, eliminating the need for a dedicated compiler. After the kernel is processed and compiled, it is linked with the Tiny-OpenCL static library to generate the final binary.
    
        In the kernel address space, a memory region is reserved for kernel arguments, which are written by the host prior to execution. These arguments include execution parameters such as the global and local sizes, pointers to input and output buffers, and other configuration details.

        To enhance flexibility, the framework allows users to customize the parameters listed in Table~\ref{tbl:gpu_knobs}. These include the base address of the kernel, which serves as the boot address of the e-GPU and enables the execution of kernels located at any address offset within the system. The location and size of thread stacks can also be configured, ensuring proper mapping and sizing according to the specific requirements of the target applications.

    \subsection{Tiny-OpenCL execution}

        After the OpenCL runtime, which is running on the host as described in Section~\ref{sec:runtime}, launches an e-GPU acceleration, kernel execution begins and proceeds through three phases: startup, scheduling, and processing.

        In the startup phase, each compute unit executes in single-thread mode, meaning that only one thread and one warp are active. During this phase, the startup functions are executed to initialize the system. This includes activating all parallel threads and concurrent warps and setting up resources such as stack pointers for each thread and warp. Once initialization is completed, compute units return to single-thread mode, and the scheduling phase begins.

        In the scheduling phase, the scheduling functions are invoked. These read the global and local sizes from the kernel arguments region to determine the number of work-groups and the number of work-items per group. These values are combined with hardware resource information (compute units, threads, and warps) retrieved from the control and status registers (CSRs). Based on this data, the scheduling functions activate the necessary resources and distribute work-items across the available threads to maximize parallelism and minimize execution time. In addition, they deactivate unused resources to reduce power consumption and improve overall energy efficiency.

        Once the work-items are launched on threads, the processing phase begins. During this phase, the user-defined kernel executes the target algorithm. Input and output parameters are fetched from the kernel arguments region, and work-items compute their specific portion of data based on their global ID, which may be expressed as a one- or two-dimensional index. Since the scheduling functions perform boundary checks in advance, the user kernel is relieved from handling such logic.

        \begin{table}[tp]
            \caption{Configurability knobs of the presented e-GPU platform.}
            \label{tbl:gpu_knobs}
            \centering
            \resizebox{0.475\textwidth}{!}{
            \begin{tabular}{|c|}
            \toprule
            \bf{Hardware}\\
            \midrule
            \textcolor{myred}{\textbullet} Number of compute units (CUs).\\
            \textcolor{mygreen}{\textbullet} Number of threads running in parallel on each CU.\\
            \textcolor{myred}{\textbullet} Number of warps running concurrently on each CU.\\
            \textcolor{mygreen}{\textbullet} Size of the instruction cache.\\
            \textcolor{myred}{\textbullet} Number of banks in the instruction cache.\\
            \textcolor{mygreen}{\textbullet} Line size of the instruction cache.\\
            \textcolor{myred}{\textbullet} Size of the data cache.\\
            \textcolor{mygreen}{\textbullet} Number of banks in the data cache.\\
            \textcolor{myred}{\textbullet} Line size of the data cache.\\
            \toprule
            \bf{Software}\\
            \midrule
            \textcolor{mygreen}{\textbullet} Base address of the kernel.\\
            \textcolor{myred}{\textbullet} Base address of the stacks.\\
            \textcolor{mygreen}{\textbullet} Size of the stacks.\\
            \bottomrule
            \end{tabular}
            }
        \end{table}

\section{APU SYSTEM} \label{sec:APU}

    This section describes how the e-GPU can be integrated with a host to realize an APU for TinyAI applications. The section first details the e-GPU interfaces and then introduces the selected host system and its configuration. Finally, it presents a lightweight host-side Tiny-OpenCL runtime developed to overcome the programming limitations inherent to resource-constrained devices.

    \subsection{e-GPU interfaces}

        The e-GPU exposes the following ports: (1)~a slave OBI port for configuration; (2)~a master OBI port for reading and writing kernel instructions and data to and from the host main memory; and (3)~an interrupt line connected to the host interrupt controller, which notifies the CPU upon kernel completion.

        The e-GPU operates within a dedicated power domain, enabling fine-grained power management. Clock-gating reduces power consumption during brief idle periods, while power-gating completely shuts down the accelerator during extended periods of inactivity. The control signals for these mechanisms are connected to the host power manager, ensuring efficient and autonomous energy management.

    \subsection{Host configuration}

        The X-HEEP~\cite{x-heep_1, x-heep_2, x-heep_3} system is selected as the platform's host due to its configurable and extensible architecture, which facilitates accelerator integration. The chosen configuration includes: (1)~a lightweight CPU~\cite{ri5cy}, optimized for executing control tasks while offloading performance-intensive computations to the accelerators, ensuring low power consumption; (2)~two SRAM banks of \SI{32}{\kibi\byte} configured in contiguous addressing mode, enabling separate memory allocation for the CPU and accelerators and reducing access conflicts; (3)~a fully connected crossbar providing high-bandwidth data transfer capabilities; (4)~an interrupt controller to manage both internal and external interrupts; (5)~a power controller responsible for managing internal and external power-saving strategies, including clock gating and power gating; (6)~a debug unit for complete system control through JTAG; and (7)~the eXtendible Accelerator InterFace (XAIF), configured with one OBI master, one OBI slave, and one interrupt interface to connect the corresponding ports of the e-GPU. Figure~\ref{fig:gpu_architecture} illustrates the details of the connection.

    \subsection{Tiny-OpenCL runtime} \label{sec:runtime}

        Standard open-source OpenCL implementations, such as PoCL~\cite{pocl}, require a host capable of running an operating system to support their runtime, which interacts with and controls the accelerators. However, X-HEEP is a RISC-V microcontroller that lacks the necessary ISA support to run an operating system like Linux. Instead, it relies on Newlib~\cite{newlib}, an open-source C standard library optimized for edge systems, which offers a complete set of basic functions, including I/O, memory, and string utilities, typically used when a full operating system with file system and multi-threading support is not available.

        Due to the absence of a file system, the host application must be compiled into a single binary, eliminating the need for dynamic library linking. Likewise, the lack of multi-threading mandates that the host application executes in a single thread.

        To overcome these limitations, the Tiny-OpenCL framework introduced in Section~\ref{sec:opencl} is extended with a dedicated runtime specifically designed for Newlib-capable microcontrollers like X-HEEP. This runtime is compiled with the standard GNU RISC-V toolchain and implements a subset of the standard OpenCL runtime API. These functions provide full control over the e-GPU from the host, including initialization of data buffers, configuration of kernel arguments, kernel dispatch, and synchronization by waiting for operations to complete.
    
\section{EXPERIMENTAL SETUP}

    This section presents the selected e-GPU configurations used for our experiments. It then introduces the adopted benchmarks, chosen to demonstrate the adaptability of the proposed platform to the tight requirements of a representative TinyAI application domain, bio-signal processing. Subsequently, it details the implementation of our systems and describes the methodology used to extract performance, power, and area metrics.

    \subsection{e-GPU configuration}

        The selected configurations are summarized in Table~\ref{tbl:config}. Each instance includes two compute units and an increasing number of parallel threads, ranging from two to eight per compute unit. This setup delivers sufficient performance for kernel execution while maintaining a balance between area and power consumption. To hide memory access latency, each compute unit is equipped with four concurrent warps. Given that the shared data cache has an access latency of four cycles, this configuration allows the cache pipeline to be filled with four outstanding requests, thus achieving a maximum throughput of one access per cycle.

        The instruction caches adopt a single-bank configuration. This arrangement is area- and power-efficient, contributing to improved overall energy efficiency. A line size of \SI{16}{\byte} (four instructions) maximizes spatial locality and enhances the pre-fetching of kernel code, while a total instruction cache size of \SI{4}{\kibi\byte} (\SI{2}{\kibi\byte} per compute unit) ensures that each kernel of our benchmarks fits.

        The shared data cache uses a multi-bank configuration with line-interleaved addressing. It features two banks, one per compute unit, to maximize parallel access efficiency. The line size is set to T x \SI{4}{\byte}, where $T$ denotes the number of parallel threads per compute unit. This design allows all threads to read data simultaneously when accessing sequential memory locations, as a single cache line fetch suffices. Finally, a total data cache size of \SI{16}{\kibi\byte} is provisioned to accommodate the required kernel data, thereby minimizing the traffic between the cache and the host main memory.

    \subsection{Benchmarks} \label{sec:kernels}
    
        Two benchmarks are employed: GeMM and TinyBio workloads. The GeMM benchmark includes generic matrix multiplication kernels of increasing sizes, ranging from $32 \times 32$ to $256 \times 256$. In contrast, the TinyBio domain includes kernels developed to analyze bio-signals acquired from the human body and extract meaningful features~\cite{bio_apps, ace}. Among these, we choose the MBio-Tracker application~\cite{mbio_tracker}, which is specifically developed to measure cognitive workload. This application features a pipeline of four stages: pre-processing, delineation, feature extraction, and prediction. These heterogeneous kernels incorporate a variety of algorithms with differing complexity and memory requirements, making them a suitable choice for our experiments.

        During pre-processing, a finite impulse response (FIR) filter is applied to the raw input data. In the delineation phase, the peaks and troughs of the filtered signal are identified to determine the inspiration and expiration times. The extracted values are then used to compute time-domain features, such as the mean, median, and root mean square (RMS), while frequency-domain features are obtained by performing a Stockham fast Fourier transform (FFT) on the filtered signal. Finally, a support vector machine (SVM) algorithm is used to estimate the cognitive workload.

    \subsection{Methodology}
    
        Each system is synthesized using TSMC’s \SI{16}{\nano\meter} FinFET SVT CMOS technology, operating at \SI{300}{\mega\hertz} and \SI{0.8}{\volt}. Post-synthesis simulations are conducted to extract the switching activity for power analysis. These data are then used to estimate energy consumption.

        \begin{table}[tp]
            \caption{Selected system configurations for our experiments.}
            \label{tbl:config}
            \centering
            \resizebox{0.475\textwidth}{!}{
            \begin{tabular}{|l|c|c|c|}
            \toprule
            \textbf{Metric} & \textbf{e-GPU 4T} & \textbf{e-GPU 8T} & \textbf{e-GPU 16T} \\
            \midrule
            Compute Units   & \multicolumn{3}{c|}{2} \\
            \midrule
            Parall. Threads & 2 xCU & 4 xCU & 8 xCU \\
            \midrule
            Concur. Warps   & \multicolumn{3}{c|}{4 xCU} \\
            \midrule
            I-Cache Size    & \multicolumn{3}{c|}{\SI{2}{\kibi\byte} xCU} \\
            \midrule
            I-Cache Banks   & \multicolumn{3}{c|}{1 xCU} \\
            \midrule
            I-Cache Line    & \multicolumn{3}{c|}{\SI{16}{\byte} xCU} \\
            \midrule
            D-Cache Size    & \multicolumn{3}{c|}{\SI{16}{\kibi\byte}} \\
            \midrule
            D-Cache Banks   & 2 & 4 & 8 \\
            \midrule
            D-Cache Line    & \SI{8}{\byte} & \SI{16}{\byte} & \SI{32}{\byte} \\
            \bottomrule
            \end{tabular}
            }
        \end{table}

        \begin{figure*}[t]
            \centering
            \includegraphics[width=0.98\textwidth]{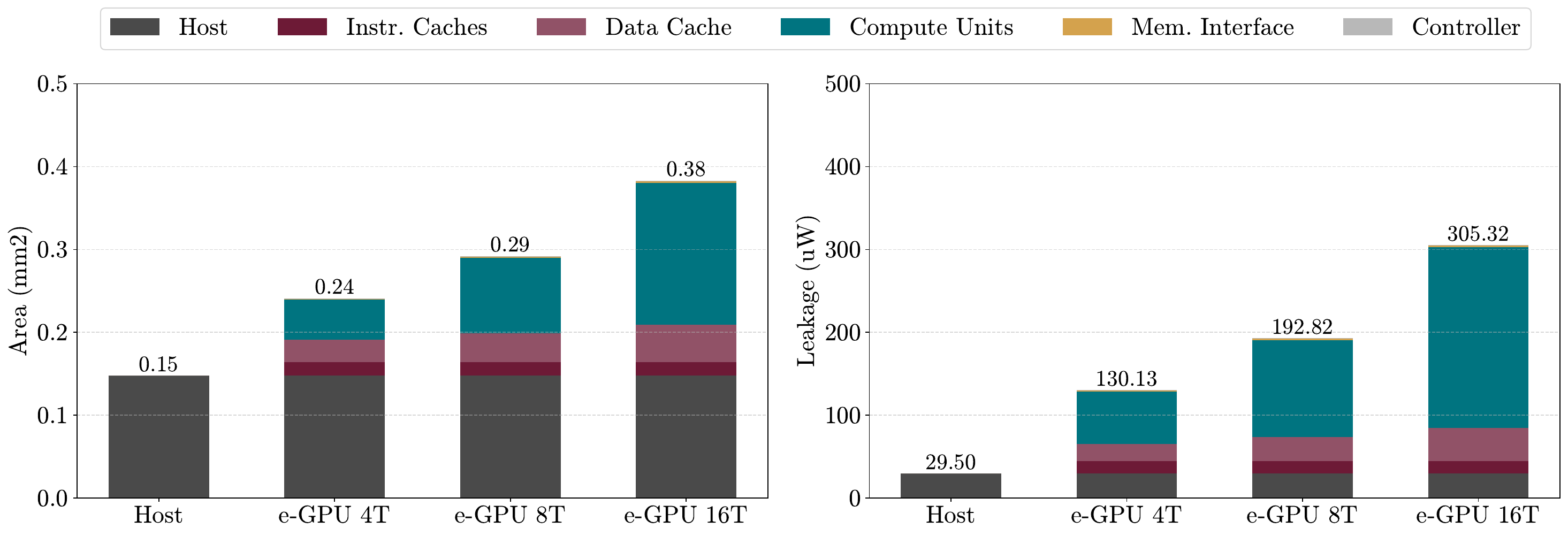}
            \caption{Breakdown of area and leakage power for the selected e-GPU systems compared to the baseline host system.}
            \label{fig:area_leakage}
        \end{figure*}

\section{EXPERIMENTAL RESULTS}

    This section presents and analyzes the results of our experiments. It begins with a static characterization of each system, followed by an examination of the overheads associated with using the e-GPU. Next, we present the characterization of the TinyBio benchmark. Finally, an overall discussion highlights the trade-offs in effectively utilizing each system.

    \subsection{Static characterization}

\begin{figure*}[t]
        \centering
        \includegraphics[width=0.98\textwidth]{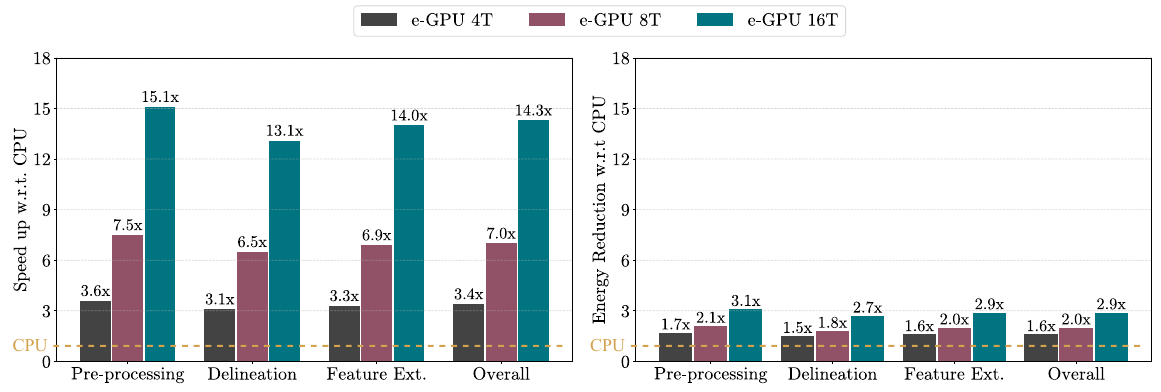}
        \caption{Speed-up and energy reduction of the TinyBio benchmark on the selected e-GPU systems compared to the baseline host system.}
        \label{fig:proc_energy}
    \end{figure*}

        Figure~\ref{fig:area_leakage} shows the breakdown of the area of the systems analyzed. The e-GPU systems occupy a total area ranging from \SI{0.24}{\milli\meter\squared} to \SI{0.38}{\milli\meter\squared}. These results demonstrate the feasibility of the proposed system for deployment in TinyAI applications, where the small form factor of target devices imposes stringent area constraints, with the entire SoC typically limited to just a few square millimeters. Furthermore, the e-GPU introduces only an area overhead of \SI{1.6}{\times} to \SI{2.5}{\times} compared to the standalone host system, which occupies \SI{0.15}{\milli\meter\squared}, while providing a speed up of up to \SI{3.6}{\times} to \SI{15.1}{\times} when running the TinyBio benchmark, as shown in Figure~\ref{fig:proc_energy}.

        Breaking down the contributions to the total area, the instruction cache remains constant across all e-GPU configurations. This behavior is expected, as the instruction cache is not modified or scaled when increasing the thread count and performance.

        In contrast, the data cache exhibits a slight increase in area, despite its total capacity remaining constant between configurations. This increase results from adaptations in the memory hierarchy aimed at maximizing memory access bandwidth under higher levels of parallelism (i.e., higher thread counts). Specifically, a higher number of cache ports (one port per thread) and larger cache lines (one word per thread) are required. The larger line size is implemented by increasing the number of banks: two banks for the 4-thread e-GPU, four banks for the 8-thread version, and eight banks for the 16-thread version. Splitting into several sub-banks is less area-efficient with respect to having a single bank of the same aggregate size.

        Finally, the area of the compute units grows significantly, and their contributions nearly double as the performance configuration scales up. This growth reflects the need to integrate additional resources, such as more arithmetic units, larger register files, and expanded control logic, to efficiently support a higher number of parallel threads.

        A similar trend is observed when analyzing the leakage power, as shown in Figure~\ref{fig:area_leakage}. The e-GPU systems exhibit leakage ranging from \SI{130.13}{\micro\watt} to \SI{305.32}{\micro\watt}, with a \SI{4.4}{\times} to \SI{10.3}{\times} overhead compared to the standalone host system, which consumes \SI{29.50}{\micro\watt}. This reinforces the suitability of the proposed design for TinyAI applications, where battery-operated devices must adhere to strict power budgets, typically in the order of tens of milliwatts.

        Overall, both area and leakage analyses consistently show that the most significant variations are localized within the compute units, while the caches remain relatively stable except for moderate adjustments in the data cache. These results highlight the scalability of the proposed e-GPU architecture, confirming its ability to efficiently trade off area and leakage overheads for substantial performance gains, shown in Figure~\ref{fig:proc_energy}, under stringent TinyAI deployment constraints.        
        
    \subsection{Overhead Characterization}

\begin{figure*}[t]
            \centering
            \includegraphics[width=0.98\textwidth]{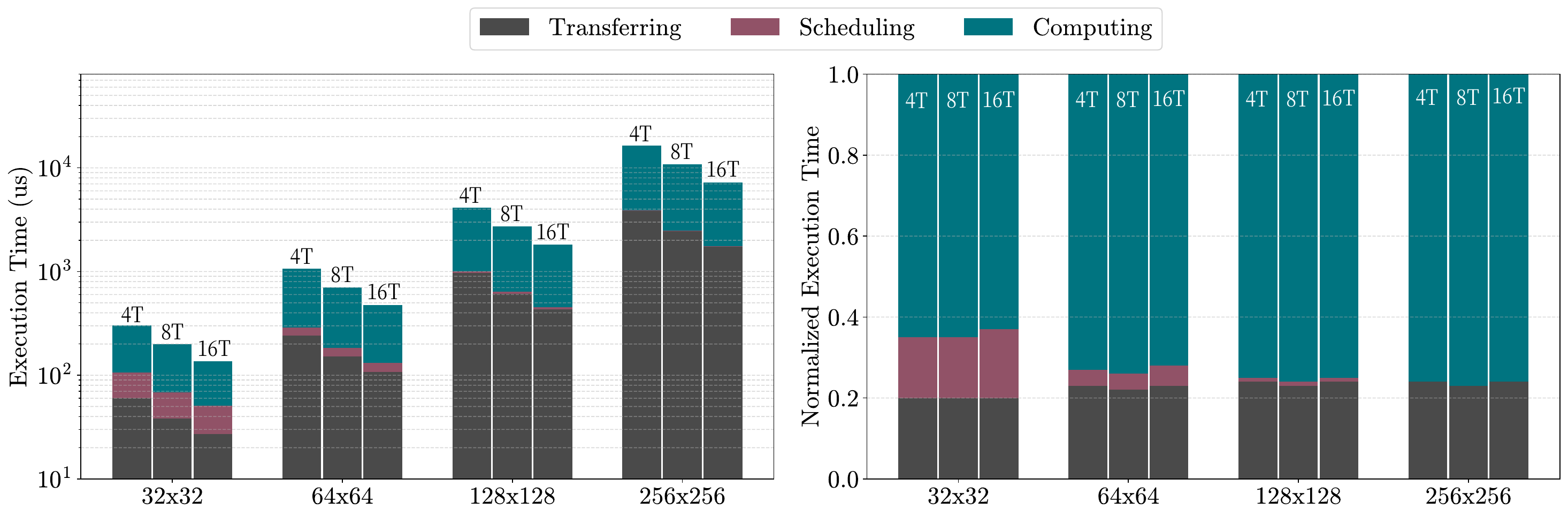}
            \caption{Breakdown of execution time (in logarithmic scale) and normalized execution time for the selected e-GPU systems (4T: e-GPU with 4 threads, 8T: e-GPU with 8 threads, 16T: e-GPU with 16 threads) across matrix multiplication kernels of increasing size.}
            \label{fig:overheads}
        \end{figure*}

        Figure~\ref{fig:overheads} shows the breakdown of the overheads versus computation time for matrix multiplications of increasing size executed on the selected e-GPU systems.

        Two types of overhead are considered: transferring and scheduling. Transferring refers to the time spent filling the data cache with kernel data. This occurs before kernel execution, as data is moved from the host memory to the e-GPU cache, with a bandwidth of \SI{32}{\bit\per\cycle}. Assuming the cache is large enough to accommodate the entire working set, the transferred data remains resident throughout the subsequent kernel executions, improving performance and energy efficiency. This transfer delay increases with matrix size, from approximately \SI{27}{\micro\second} to \SI{1.7}{\milli\second} for the high-range e-GPU system, since progressively more data must be transferred.

        At a fixed matrix size, increasing performance using more parallel threads results in a slight decrease in transfer time. This effect can be attributed to progressively larger cache lines, which enable more data to be pre-fetched into the data cache with each read miss. As a result, the overlap between computation and data transfer increases, reducing the fraction of time spent solely on data movement.

        Overall, the transferring overhead stabilizes at slightly more than \SI{20}{\percent} of the execution time across matrix sizes, as the increase in transfer time is compensated by the reduction in computation delay due to progressively higher performance.

        Complementary to transferring, scheduling represents the time required to allocate computing resources and assign work-items to them. The e-GPU relies on run-time scheduling provided by the Tiny-OpenCL framework described in Section~\ref{sec:opencl}. This overhead becomes more pronounced as the number of work-items increases, since more iterations over the available resources must be performed.

        In our experiments, we configured the matrix multiplication kernels so that the number of work-items matches the number of available threads (compute units × concurrent warps × parallel threads). This approach optimizes scheduling time and improves overall kernel performance. As a result, scheduling time remains constant at approximately \SI{25}{\micro\second}, regardless of increases in system performance or matrix size.

        However, its relative contribution to the total execution time decreases, from approximately \SI{15}{\percent} to less than \SI{1}{\percent}, as the size of the problem increases. This demonstrates the feasibility of the presented Tiny-OpenCL framework, which offers high flexibility and improved programmability with negligible overhead for matrix sizes larger than $256 \times 256$ (or equivalent problem sizes).

    \subsection{TinyBio Benchmark Characterization}
    
        Figure~\ref{fig:proc_energy} presents the speed up and energy reduction achieved by each kernel of the TinyBio benchmark, described in Section~\ref{sec:kernels}, when executed on the analyzed e-GPU systems, relative to the baseline host system. Since the impact of each overhead has been analyzed previously, transfer and scheduling overheads are excluded from these experiments, and only the pure computation phase is considered. Consequently, the data cache is assumed to be pre-populated with the input data required by each kernel, and a static kernel scheduling policy is employed.
        
        During the pre-processing phase, the FIR filter is executed. This algorithm is highly parallelizable, as each output sample can be computed independently, allowing multiple threads to process different samples concurrently. The computation involves a fixed number of multiply-accumulate operations, which efficiently map to the processing units of the e-GPU. Moreover, the FIR filter exhibits regular and sequential memory access patterns, enhancing locality and enabling efficient coalesced memory accesses, thereby minimizing latency and maximizing bandwidth utilization.
        
        Thanks to these favorable characteristics, the e-GPU systems achieve a speed up ranging from \SI{3.6}{\times} to \SI{15.1}{\times} over the host system, resulting in a \SI{1.7}{\times} to \SI{3.1}{\times} reduction in energy consumption.
        
        In contrast, the delineation phase is characterized by a control-intensive algorithm used to detect local maxima and minima. This workload is only partially parallelizable and does not fully exploit the capabilities of the e-GPU. Furthermore, GPUs are not optimized for control-dominated codes involving frequent branching, where divergent execution paths are serialized using thread masking, thereby reducing parallel efficiency.
        
        Despite these limitations, the e-GPU systems achieve a \SI{3.1}{\times} to \SI{13.1}{\times} speed up over the host, with an associated \SI{1.5}{\times} to \SI{2.7}{\times} reduction in energy consumption.
        
        During the feature extraction phase, the Stockham FFT algorithm is executed. Unlike the traditional Cooley–Tukey FFT, the Stockham FFT eliminates the need for a separate bit-reversal permutation by reordering data at each stage. It employs a double-buffering (ping-pong) scheme, alternating between two arrays to store intermediate results. At each of its $\log_2(N)$ stages, it performs independent butterfly operations with regular and sequential memory accesses, making it particularly well-suited for parallel architectures. By the end of the computation, the output data is already in the correct order, eliminating the need for an explicit final permutation step.
        
        However, the main limitation of this stage is the requirement for synchronization between sequential stages, which reduces available parallelism and constrains overall performance. Nevertheless, the e-GPU systems deliver a \SI{3.3}{\times} to \SI{14.0}{\times} speed up, accompanied by a \SI{1.6}{\times} to \SI{2.9}{\times} reduction in energy consumption.

        The SVM is executed after the feature extraction stage to perform final classification. Given its negligible execution time, it is included within the feature extraction phase.

        Overall, the e-GPU systems achieve an average performance improvement ranging from \SI{3.4}{\times} to \SI{14.3}{\times} during the processing phases, along with an energy consumption reduction ranging from \SI{1.6}{\times} to \SI{2.9}{\times}.

    \subsection{Overall discussion}

        The characteristics of the evaluated systems highlight critical trade-offs that must be considered when designing or selecting platforms for TinyAI applications.

        The baseline host system offers the lowest complexity, making it well-suited for scenarios with modest computational demands and stringent resource constraints. Its lightweight design ensures minimal area and power usage, which is crucial for applications where energy availability is severely limited, such as battery-powered edge devices. However, these benefits come at the expense of computational capability. The host CPU struggles to efficiently handle the demands of more intensive TinyAI workloads, leading to prolonged execution times and, consequently, higher overall energy consumption. As TinyAI applications increasingly require greater processing power to support more complex models and data analytics, the baseline host CPU becomes less viable for such use cases.

        In contrast, the e-GPU systems introduce moderate overhead in both area and leakage compared to the host system. However, these additional costs are offset by significant gains in performance and energy efficiency. The e-GPU delivers speed ups ranging from \SI{3.6}{\times} to \SI{15.1}{\times} and achieves energy savings of \SI{1.7}{\times} to \SI{3.1}{\times}. These improvements are particularly valuable for TinyAI applications that involve intensive data processing but must still operate within tight area and power budgets. Although control-intensive workloads remain more challenging for the e-GPU, due to the intrinsic inefficiencies of SIMT architectures when managing divergent branches, the system handles such cases effectively enough to sustain performance and energy efficiency.

        Overall, the e-GPU system emerges as a highly effective and versatile solution for TinyAI applications. Its architecture successfully balances computational throughput, energy efficiency, and flexibility, addressing the key challenges faced in resource-constrained environments. By significantly improving performance and reducing energy consumption while maintaining a modest area and power footprint, the e-GPU meets the stringent requirements of modern TinyAI workloads.

\section{CONCLUSION}

    The growing demand for real-time, machine-learning-based computing has accelerated the adoption of edge computing, where local data processing enhances latency, privacy, and energy efficiency. However, these workloads impose strict constraints on computational performance, area, and power consumption, necessitating specialized hardware solutions. Among available accelerators, GPUs offer strong potential in leveraging data parallelism for machine learning and signal processing tasks. Yet, their trade-offs in the context of TinyAI devices were largely unexplored.

    To address these challenges, this work has introduced e-GPU, an open-source and configurable RISC-V GPU platform designed specifically for TinyAI applications. Its extensive configurability enables fine-grained area and power optimizations, while the dedicated Tiny-OpenCL framework provides a lightweight and flexible programming model tailored for resource-constrained environments.

    The e-GPU is integrated with the X-HEEP host~\cite{x-heep} to realize an APU optimized for TinyAI applications. Multiple instances of the proposed system, featuring varying e-GPU configurations, are implemented in TSMC’s \SI{16}{\nano\meter} SVT CMOS technology and are operated at \SI{300}{\mega\hertz} and \SI{0.8}{\volt}. Their area and leakage characteristics are analyzed to ensure alignment with TinyAI constraints. To assess both runtime overheads and computational efficiency, we employ two benchmarks: GeMM and TinyBio workloads. The GeMM benchmark is used to quantify the scheduling overhead introduced by the Tiny-OpenCL framework. The results showed that the delay becomes negligible for matrix sizes larger than $256 \times 256$ (or equivalent problem sizes). The TinyBio benchmark is then used to evaluate performance and energy improvements over the baseline host under pure processing conditions. The results indicate that the high-range e-GPU configuration with 16 threads achieves up to a \SI{15.1}{\times} speed-up and reduces energy consumption by up to \SI{3.1}{\times}, while incurring only a \SI{2.5}{\times} area overhead and operating within a \SI{28}{\milli\watt} power budget.

    By providing an open-source and highly configurable GPU platform with competitive area, power, and programmability, this work lays the foundation for further research into GPU platforms for TinyAI applications, opening new opportunities for flexible and energy-efficient computing at the edge.

\section*{Acknowledgments}

    We would like to thank the entire \mbox{X-HEEP} team for their significant contributions to the platform. In particular, we express our gratitude to Dr. Miguel Peón-Quirós for his valuable support in the architectural design of the system.

\bibliography{references}

\end{document}